\documentstyle[pre,aps,epsf,psfig]{revtex}
\newcommand \beq{\begin{eqnarray}}
\newcommand \eeq{\end{eqnarray}}
\newcommand \be{\begin{eqnarray}}
\newcommand \ee{\end{eqnarray}}
\newcommand{\set}[2]{\newcommand{#1}{#2}}
\set{\pa}{\partial \over \partial\, }
\set{\leftvector}{\stackrel{\leftarrow}{\partial }}
\set{\rightvector}{\stackrel{\rightarrow}{\partial }}
\begin{document}
\twocolumn[\hsize\textwidth\columnwidth\hsize
           \csname @twocolumnfalse\endcsname
\title{
Nonlinear relaxation field in charged
systems under high electric fields}
\author{K. Morawetz}
\address{
LPC-ISMRA, Bld Marechal Juin, 14050 Caen
 and  GANIL,
Bld Becquerel, 14076 Caen Cedex 5, France
}
\maketitle
\begin{abstract}
The influence of an external electric field on the current in charged
systems is investigated. The results from the classical
hierarchy of density matrices are compared with the results from the quantum
kinetic theory. The kinetic theory yields a systematic
treatment of the nonlinear current beyond linear response. To this end
the dynamically
screened and field-dependent Lenard-Balescu equation is integrated
analytically and the nonlinear relaxation field is calculated. The 
classical linear response result known as Debye - Onsager
relaxation effect is only obtained if asymmetric screening is
assumed. Considering the kinetic equation of one specie the
other species have to be
screened dynamically while the screening with the same specie
itself has to be performed statically. Different other approximations are
discussed and compared.
\end{abstract}
\vskip2pc]

\newcommand{\grlo}{\stackrel{>}{<}}
\newcommand{\logr}{\stackrel{<}{>}}

\section{Introduction}

High field transport has become a topic of current interest in various fields of physics. In semiconductors the nonlinear
transport effects are accessible due to femto - second laser pulses and shrink devices \cite{HJ96}. In plasma physics these
field effects can be studied within such short pulse periods \cite{THWS96}. One observable of interest is the current or
the electrical conductivity which gives access to properties of dense nonideal plasmas \cite{kker86}. In high energy
physics the transport in strong electric fields is of interest due to pair creation \cite{KES93}.
In order to describe these field effects one can start conveniently from kinetic theory. Within this approach the crucial question is to derive appropriate kinetic equations which include field effects beyond linear response. 

At low strength of the external electric field one expects the linear response regime to be valid. Then the contribution of field effects to the conductivity
can be condensed into the Debye- Onsager relaxation effect \cite{k58,KE72,e76,er79,r88,MK92,ER98} which was first derived within
the theory of electrolytes \cite{DH23,O27,f53,FEK71,KKE66}. Debye has given a limiting law of electrical conductivity
\cite{DH23} which stated that the external electric field $E$ on a
single charge $Z=1$ is diminished in an electrolyte solution by
the amount
\beq\label{debye}
\delta E=E \,(1- {\kappa e^2 \over 6 T})
\eeq
where  $e$ is the elementary charge, $E$ the electric field strength,  $T$
is the temperature of the plasma and $\kappa$ is the inverse screening
radius of the screening cloud. This law is interpreted as a
deceleration force which is caused by the deformed screening cloud
surrounding the charge. Later it has been shown by Onsager
\cite{O27} that this
result has to be corrected to
\beq\label{ons}
\delta E=E \,(1- {\kappa e^2  \over 3 (2 +\sqrt{2})T})
\eeq
if the dynamics of ions ($Z=1$) is considered. 
While the linear response theory seems to reproduce this Onsager result \cite{er79,r88,ER98},
the kinetic theory seems to support more the Debye result \cite{MK92,Ma97,ER98}. The
correct treatment is a matter of ongoing debate. In this paper we will
give the result beyond linear response for the statically and
dynamically screened approximation. Here different approximations of kinetic theory 
will be discussed and the one which leads to the closest form to the hydrodynamical approximation
(Onsager result) is presented.

The kinetic approach describes the time evolution of
the one particle
distribution function within an external field ${\bf E}$ as
\be
{\partial \over \partial t} f- e Z\, {\bf E}{\partial \over \partial
  {\bf k}} f=
I[f,{\bf E}]
\label{ek}
\ee
where the field dependent collision integral $I[f,{\bf E}]$ has to be provided by
different approximations. Integrating this kinetic equation over the
momentum ${\bf k}$ one obtains the balance of the current. For
simplicity we assume that the distribution function can be
parameterized by a displaced local equilibrium one with a field and
time dependent momentum $f({\bf k},t)=f_0({\bf k}-{\bf p}({\bf
  E},t))$ which is related to the current ${\bf J}$ as
\be
{\bf J}(E)=n Z e {{\bf p}(E)\over m}
\ee
if the charge is $Z e$, the density $n$  and the mass $m$. The balance equation for the
field and time dependent local momentum ${\bf p}(E,t)$
follows from (\ref{ek}) as
\be
{\partial \over \partial t} {\bf p}-e Z n  (1+{\delta E(E)\over E})\,{\bf
  E}=R(E) e Z n\,  {\bf J}
\label{ek1}
\ee
where the relaxation field $\delta E(E)$ as well as the free
conductivity $R(E)$ follows from the field dependent collision
integral. The total conductivity  ${\bf E}=\sigma {\bf J}$ is then
given by
\be
\sigma (E)={R(E)\over 1+{\delta E(E)\over E}}.
\ee
The free conductivity $R$ is the subject of intense investigations in
the literature \cite{kker86}. It is known that the Coulomb divergence for small wave vectors
is omitted if screening is included and the divergence at large
wave vectors is omitted by the De Broglie wavelength i.e. by the quantum
effects. We will not consider the discussion of the free conductivity
$R$ here but concentrate on the relaxation field $\delta E$. The free
conductivity can be obtained by the same considerations as will be
outlined here. We want to point out that the relaxation field will
turn out to be free of long wave divergences in the classical limit in contrast
to the free conductivity $R$.

First we recall the  hydrodynamical approach starting from the
classical Bogoliubov-Born-Green-Kirkwood-Yvon (BBGKY) hierarchy which results into an analytical formula for
the classical relaxation effect already reported \cite{O97}.
This result is then compared with the quantum kinetic approach.
We give a short rederivation of the field dependent kinetic equations
in dynamical screened approximation from the Green's function technique in Sec.~III. Two approximations, the static screening as well as dynamical screening are presented.
In the fourth
section we will derive the field dependent current analytically.
We present both the statically as well as dynamically screened
treatment as analytical results. The classical expressions for the
statically screened result \cite{Ma97} is compared with the classical
result from hydrodynamical approximation.
The dynamical result is then derived analytically too and it will be
shown that only for asymmetric screening the hydrodynamical result can
be approached. In Sec.~V we shortly discuss the physical limitation of
field strengths
for the local equilibrium assumption and the gradient
approximation. Sec.~VI summarizes and the appendix gives the
calculation of some involved integrals appearing during the
integration of the Lenard-Balescu equation.

\section{Approach by classical BBGKY-Hierarchy} \label{Hydrodyn}

The starting point for the classical considerations is the
BBGKY hierarchy \cite{B46,KE62} which reads for the
one - particle distribution function $F_a$ 
\begin{eqnarray}\label{2bbgky}
&&\frac{\partial{F_a}}{\partial t} \, + {\bf v} \,
\frac{\partial{F_a}}{\partial{{\bf r}}} + \frac{e_a}{m_a} {\bar {\bf E}} \frac{\partial{F_a}}{\partial{{\bf v}}} - S_a F_a \nonumber \\
&&= \sum_b \! \frac{n_b e_a e_b}{m_a} \frac{\partial{}}{\partial {\bf
    v}} \! \int \!\! d {\bf r'} \, d {\bf v'} 
F_{ab}({\bf r},{\bf r'},{\bf v},{\bf v'}) \frac{\partial{}}{\partial {\bf
    r}} \, \frac{1}{\left| {\bf r} - {\bf r'} \right|}\nonumber\\
&&
\end{eqnarray}
and the two - particle
distribution function $F_{ab}$
\begin{eqnarray}\label{2bbgky2}
&&\frac{\partial{F_{ab}}}{\partial t} \, + {\bf v} \,
\frac{\partial{F_{ab}}}{\partial{{\bf r}}} +  {\bf v'} \,
\frac{\partial{F_{ab}}}{\partial{{\bf r'}}} + \frac{e_a}{m_a} {\bar {\bf E}} \frac{\partial{F_{ab}}}{\partial{{\bf v}}} \nonumber \\
&&+  \frac{e_b}{m_b} {\bar {\bf E}} \frac{\partial{F_{ab}}}{\partial{{\bf v'}}}
- S_a F_{ab} - S_b F_{ab} \nonumber \\
&&= e_a e_b \frac{\partial{}}{\partial {\bf r}} \, \frac{1}{\left| {\bf r} - {\bf r'} \right|} \left( \frac{1}{m_a}  \frac{\partial{F_{ab}}}{\partial{{\bf v}}}- \frac{1}{m_b}  \frac{\partial{F_{ab}}}{\partial{{\bf v'}}} \right) \nonumber \\
&&+ \sum_c {n_c e_c}  \int d {\bf r''} \, d {\bf v''}  \left (
{e_a\over m_a} \frac{\partial{}}{\partial {\bf r}}
  \, \frac{1}{\left| {\bf r} - {\bf r''} \right|} \cdot {\partial
    F_{abc} \over \partial {\bf v}}\right .
\nonumber\\
&&\left .+
{e_b\over m_b} \frac{\partial{}}{\partial {\bf r'}}
  \, \frac{1}{\left| {\bf r'} - {\bf r''} \right|} \cdot {\partial
    F_{abc} \over \partial {\bf v'}}
\right ) 
\end{eqnarray}
with the external field ${\bf E}$. $S_a$ describes a collision integral
with some background which we will specify later.
This hierarchy is truncated  approximating that \cite{KE62} 
\begin{eqnarray}\label{weakcoupl}
F_{ab} &=& F_a F_b + g_{ab} \nonumber\\ 
F_{abc} &=& F_a F_b F_c + F_a \,g_{bc} + F_b \,g_{ac} + F_c \,g_{ab}
\end{eqnarray}
where $g_{ab}({\bf r_a},{\bf r_b},{\bf v_a},{\bf v_b})$ is the two-particle correlation function.

Within the local equilibrium approximation we suppose a
stationary (for example a local Maxwellian) distribution for the velocities in
the one and two-particle distribution functions
\begin{eqnarray}\label{Maxwell}
&&f_a({\bf r},{\bf v},t) = n_a({\bf r},t)\left( \frac{m_a}{2 \pi T}
\right)^{3/2} \exp{\left[ - \frac{m_a ({\bf v} - {\bf u_a})^2}{2T} \right]}
\nonumber \\
&&g_{ab}({\bf r,r',v,v'},t) = F_{ab}-F_a F_b\nonumber\\
&&=h_{ab}({\bf r,r'},t)\left( \frac{m_a m_b}{4 \pi^2 T^2} \right)^{3/2}\nonumber\\&&
\times\exp{\left[ - \frac{m_a ({\bf v - w_{ab}})^2}{2T}  - \frac{m_b ({\bf v' - w_{ba}})^2}{2T} \right]}.
\end{eqnarray}
Here we have introduced the local one-particle density and the local average velocity
\be
n_a({\bf r},t) &=& \int d {\bf v} F_a({\bf r},{\bf v},t) \, \nonumber\\
{\bf u_a} &=&  \frac{1}{n_a} \int d {\bf v} {\bf v} \, F_a({\bf r},{\bf v},t)
\ee
as well as the pair correlation function and the average pair velocity
\be
h_{ab}({\bf r},{\bf r'},t) &=& \int d {\bf v} \, d {\bf v'} g_{ab}({\bf r},{\bf r'},{\bf v},{\bf v'},t)  \, ,\nonumber\\
\quad w_{ab}({\bf r},{\bf r'},t) &=&   \frac{1}{h_{ab}} \int d {\bf v}
\,  d {\bf v'}{\bf v} g_{ab}({\bf r},{\bf r'},{\bf v},{\bf v'},t).
\nonumber\\&&
\ee
Further on, we suppose that the particles interact with some
background (e.g. neutrals or electrolyte solvent) by the collision integrals $S_a$ with the following properties
\begin{eqnarray}\label{Stoss}
\int d {\bf v} S_a f_a  &=& 0  \nonumber \\
\int d {\bf v} {\bf v} S_a f_a  &=&  \frac{1}{b_a m_a}  \rho_a {\bf u_a} \quad , \nonumber \\
\int d {\bf v} {\bf v} S_a g_{ab}({\bf r,r',v,v'},t)  &=&
\frac{1}{b_a m_a}  h_{ab} {\bf w_{ab}}
\end{eqnarray}
where $b_a$ is the mobility of particle of type a. This friction with
a background serves here to couple the two - particle equations and will be 
considered infinitesimal small in the end. However, as we will
demonstrate this yields to a symmetry breaking in the system which
leads basically to different results than neglecting this friction.

Fourier transform of the resulting two equations (\ref{weakcoupl}) into momentum space and
assuming a homogeneous density $n({\bf r})=n$ we arrive at the coupled
equation system
\be
&&-{e_a\over T} {\bar {\bf E}}. ({\bf v_a}-{{\bf u_a}}) f_a={{{\bf u_a}}\over b_a m_a}+\sum\limits_b {4 \pi
 \, n_b \,e_a e_b\over T}  \nonumber\\
&&\times\int {d{\bf k}\over (2 \pi)^3} {i {\bf k}.({\bf
    v_a-w_{ab}})\over k^2} f_a({\bf v_a-w_{ab}+u_a}) h_{ab}({\bf k})
\nonumber\\
&&
\ee
and
\be
&&i {\bf k}.({\bf v_a\!-\!v_b}) g_{ab}
\!-\!(e_a({\bf v_a\!-\!w_{ab}})\!+\!e_b ({\bf v_b\!-\!w_{ba}})){{\bar {\bf E}} \over T}g_{ab}
\nonumber\\
&&=
-ie_a e_b {4 \pi \over k^2} {\bf k}.({\bf v_a}\!-\!{{\bf u_a}}\!-\!{\bf v_b}\!+\!{{\bf u_b}}) f_a f_b\nonumber\\
&&-i\int\!\! {d {\bar {\bf k}} \over (2 \pi)^3} 
{4 \pi e_a e_b\over T k^2} 
{\bf {\bar k}}.({\bf v_a\!-\!w_{ab}\!-\!v_b\!+\!w_{ba}})g_{ab}({\bf k}\!-\!{\bar {\bf k}})\nonumber\\
&&-\sum\limits_c n_c \int d {\bf v_c} {4 \pi i e_c \over T k^2}\left
  [e_a {\bf k}.({\bf v_a}-{{\bf u_a}}) \,f_a \,g_{cb}({\bf k})\right .
\nonumber\\
&&\left .-e_b {\bf k}.({\bf v_b}-{ {\bf u_b}}) \,f_b \,g_{ac}({\bf k})\right ]+S_a g_{ab} + S_b g_{ab}
\label{hier}
\ee 
with
\be
{\bar {\bf E}}={\bf E}-\sum\limits_b n_b e_b \int d {\bf r_b} d {\bf v_b} {\partial \over \partial {\bf r_b}} {1\over |{\bf r_a-r_b}|} F_b.
\ee
By multiplying the above equation system by $1,{\bf v_a,v_b}$ and
integrating over the velocities we obtain the Onsager equation \cite{O27}
\be
&&b_a \left [T h_{ab}({\bf k}) (1+i {e_a\over e} a)+e_a \Phi_b(-{\bf k})\right ]
\nonumber\\
&&=-b_b \left [T h_{ab}({\bf k}) (1-i {e_b\over e} a)+e_b \Phi_a({\bf
    k})\right ]
\label{ons2}
\ee
with
\be
k^2 \Phi_a({\bf k})&=&4 \pi e_a +\sum\limits_c n_c e_c h_{ac}({\bf k})\nonumber\\
k^2 \Phi_a(-{\bf k})&=&4 \pi e_a +\sum\limits_c n_c e_c h_{ca}({\bf k})
\label{ons1}
\ee
for the two -particle correlation function $h_{ab}$. Here we use 
\be
a={e {{\bf k.{\bar E}}} \over k^2 T}.
\ee 

Let us already remark here that the friction with a background
described by the mobilities $b$ couple the two sides of the equation
(\ref{ons2}). If we had not considered this friction, $S_i=0$, we would have obtained that
the left and the right hand side of (\ref{ons2}) vanish
separately. This will lead essentially to a different result even for
infinite small friction. There is no continuous transition between
these two extreme cases pointing to a symmetry breaking in the two treatments. 
Let us first discuss the case with background friction.

\subsection{With background friction}

The system (\ref{ons2}) for electrons, $e_e=e$, and ions, $e_i=-Ze$,
with charge $Z$ reads expanded
\be\label{sys}
T h_{ee} &=&-e
{\Phi_e(-{\bf k})+\Phi_e({\bf k}) \over 2} \nonumber\\
T h_{ei} &=&-e {\Phi_i(-{\bf k})-Z {b_i\over b_e} \Phi_e({\bf k})\over 1+{b_i\over b_e}+i a(1+{b_i\over b_e} Z)}\nonumber\\
T h_{ie} &=&-e {\Phi_i({\bf k})-Z {b_i\over b_e} \Phi_e(-{\bf k})\over 1+{b_i\over b_e}-i a(1+{b_i\over b_e} Z)}\nonumber\\
T h_{ii} &=&Z e
{\Phi_i(-{\bf k})+\Phi_i({\bf k}) \over 2}.
\ee
This we can solve together with (\ref{ons1}). First we calculate the effective field
strength at the position of the electron in linear response the
Onsager result \cite{O27}
\be
{\delta E\over E}{\bf E}&=&-i {{\bf E}\over E}{1\over(2 \pi)^2}
\int\limits_0^{\infty} k^3 dk \int\limits_{-1}^1d(\cos \theta) \cos
\theta \Phi_e({{\bf k}})\nonumber\\
&=&{\bf E} {\kappa e^2\over 3 T} {Z q\over{\sqrt{q}+1}}
\label{erg1}
\ee
with $\kappa^2=\kappa_e^2(1+Z)={4\pi (e^2 n_e+Z^2 e^2 n_i)\over T}$ and
\be
q={b_e+Z b_i\over (1+Z)(b_e+b_i)}.
\label{erg1a}
\ee
For single charged ions $Z=1$ the influence of the mobilities drop out
and we recover the result (\ref{ons}). 

Since this result is
independent of the mobilities one could conclude that this is an
universal limiting law. However we will express two doubts here. As
one sees for charges $Z>1$ the result (\ref{ons}) is approached only in
the limit where the ion mobilities are much smaller than the electron
mobilities $b_i/b_e\rightarrow 0$. This means of course that the
electrons have different friction with a thought background than the
ions. In other words there is an explicit symmetry breaking mechanism
included by assuming such collision integrals with the
background. Therefore we will obtain another solution if we consider
no friction.

The second remark concerns the limit of one-component system which one
can obtain by setting $Z=-1$. The Onsager result or hydrodynamical
result with friction (\ref{erg1}) leads to twice the Debye result
(\ref{debye})  in this
case but with opposite sign. Oppositely we will see in the following
that the perfectly symmetric treatment of the species without friction
with a background will lead to a vanishing one component limit as it
should. 
This
again underlies the symmetry breaking if one assumes an infinitesimal
small friction with a background.

For completeness, we want to recall the expression of the nonlinear
Onsager result \cite{O97,Ma97} which is obtained from the limit
$b_i/b_e\rightarrow 0$ of the system (\ref{sys}) 
\beq\label{sys1}
T h_{ee} +e
{\varphi_e(-{\bf k})+\varphi_e({\bf k}) \over 2} &=& 0 \nonumber\\
h_{ei} (T+i \, e {{\bf k}  {\bf E} \over k^2 })+e
\varphi_i(-{\bf k})&=&o({b_i\over b_e})=0\nonumber\\
-h_{ie} (T-i \, e {{\bf k}  {\bf E} \over k^2})-e
\varphi_i({\bf k})&=&o({b_i\over b_e})=0\nonumber\\
T h_{ii} -Z e
{\varphi_i(-{\bf k})+\tilde \varphi_i({\bf k}) \over 2}&=&0.\nonumber\\
\eeq
One obtains \cite{f53,O97} the result for $Z=1$
\begin{eqnarray} \label{9}
\delta {\bf E}&=& - \frac { e^2 \kappa_e}{3(1+\sqrt{2}) \, T} {\bf
  E}\, F_H({eE\over T\kappa_e})\nonumber\\
&=&- \frac { e^2 \kappa_e}{6 \, T} {\bf E}
\left\{
\matrix{ 2-\sqrt{2}+o(E)\cr\cr
{3 \kappa T\over 2 e E}+o(1/E)^2}
\right .
\ee
with
\be
F_H(\alpha) &=& \frac{3(1+\sqrt{2})}{\alpha^2} \left[ \frac{1}{2}
  \sqrt{\alpha^2 \!+\! 2} \!-\! 1 \!+\! \frac{1}{\alpha}\arctan(\alpha) \right .
\nonumber\\
&&\left .- \frac{1}{\alpha}\arctan(\frac{\alpha}{\sqrt{\alpha^2 + 2}})\right]
.
\end{eqnarray}
The numerical values of this result will be discussed in chapter (\ref{dis}).

\subsection{Without background}

Now we reconsider the steps from (\ref{hier}) to (\ref{ons2}) without
friction with the background. We obtain that both sides of (\ref{ons2})
vanish separately
\be
&&T h_{ab}({\bf k}) (1+i {e_a\over e} a)+e_a \Phi_b(-{\bf k})=0\nonumber\\
&&T h_{ab}({\bf k}) (1-i {e_b\over e} a)+e_b \Phi_a({\bf k})=0.
\label{ons3}
\ee
Both equations have identical solutions $h_{ab}$ which can be easily
verified using the symmetry $h_{ab}({\bf k})=h_{ba}(-{\bf k})$. 
Together with (\ref{ons1}) we can solve for $\Phi_e$
and the relaxation field is obtained instead of (\ref{9})
\be
\delta {\bf E}&=&-{e^2\kappa_e \sqrt{1+Z} \over 6 T} (Z+1) {\bf
  E}F_N({e E\over T\kappa_e})
\ee
which takes for $Z=1$
\be
{\delta E\over E}=- \frac { e^2 \kappa_e}{6 \, T} 
\left\{
\matrix{ 2+o(E)\cr\cr
{3 \kappa T\over  e E}+o(1/E)^2}
\right .
\label{erg2}
\ee
with
\be
&&F_N(\alpha)={3\over (1+Z)\alpha^2}\left
[\sqrt{4+(1+Z)\alpha^2}
\right .
\nonumber\\
&&\left .
+{4\over
\sqrt{1+Z}\alpha}\log{{2\over
\sqrt{1+Z}\alpha+\sqrt{4+(1+Z)\alpha^2}}}\right].
\ee
We see that the linear
response result for $Z=1$ is twice the Debye result (\ref{debye}). 
For equal charged
system $Z=-1$ which would coincide with a one component plasma no
relaxation effect appears as one would expect. In other words in a perfectly
symmetric mathematical two - component plasma there is another
relaxation effect than in a system which distinguishes the components by
a different treatment of friction. The Onsager result (\ref{9}) does
not vanish for the limit of one-component plasma $Z=-1$. This is due
to the different treatment of ions and electrons there which assumes 
explicitly a two component plasma. Therefore the limit $Z=-1$ does not
work there.

This result is quite astonishing. One would expect that the limiting
procedure which transforms the system (\ref{ons1}) into (\ref{ons3})
would also lead to a smooth transitions of the end results. However
this is not the case. While the separate limit of $b_{e,i}\rightarrow
\infty$ of (\ref{ons1}) leads to (\ref{ons3}) there is no possibility
to transform the result (\ref{erg1}) into the linear response result
of (\ref{erg2}). This
underlines that due to even infinitesimal small friction assumed in
obtaining (\ref{erg1}) there occurs a symmetry breaking in the
sense that the electrons and ions are not anymore symmetrically
treated. 

This lesson we have to keep in mind when we now advance and
investigate the systematic treatment by quantum kinetic theory. There
we will find also complete different results when we use asymmetric
screening compared to symmetric screening. Of course, we will not assume any phenomenological
friction since the kinetic theory provides for a systematic
description of all occurring processes. Here we want only to point out
that the above symmetry breaking is the main reason for the confusion
in literature. Following the linear response formalism an asymmetric treatment of two - particle
correlation functions is used in that the electrons are statically screened
\cite{ER98}. This seemingly innocent usage leads there to an occasional agreement for
$Z=1$ with the Onsager result (\ref{ons}).

Another advantage of the kinetic theory we want to point out here. The
classical local equilibrium or hydrodynamical approximation does not lead to a mass
dependence of the relaxation effect. This will be provided by the
kinetic theory.

\section{Quantum kinetic theory} \label{greens}

We will formulate the kinetic theory within gauge invariant functions not missing field effects.
The most promising
theoretical tool is the Green function technique \cite{HJ96,D84,SL94,SL95}.
The resulting equations show some typical deviations from the ordinary Boltzmann equation: (i) A collision broadening which
consists in a smearing out of the elementary energy conservation of scattering. This is necessary to ensure global energy
conservation \cite{M94}. (ii) The intra-collisional field effect, which gives additional retardation effects in the momentum
of the distribution functions. This comes mainly from the gauge invariance.

One of the most important questions is the range of applicability of these kinetic equations. Up to which field strengths
are such modifications important and appropriate described within one-particle equations? In \cite{lkaw91} this question
has been investigated for semiconductor transport. It was found that for
high external fields the
intra-collisional field effect becomes negligible.
This range is given by a
characteristic time scale of field effects $\tau_F^2=m \hbar /(e {\bf E \cdot q})$
which has to be compared with the inverse collision frequency.
This criterion is a pure quantum one. It remains the question whether there are also criteria in the classical limit. For a
plasma system we will discuss in Sec.~V that there is indeed a critical value of the field strength which can be given by classical
considerations.

\subsection{Definitions}

In order to describe correlations in highly nonequilibrium
situations, we define
various correlation functions by different products of creation and annihilation
operators

\begin{eqnarray}\label{correlation}
G^> (1,2) & = & <\Psi(1)\Psi^+(2)> \nonumber\\
G^< (1,2) & = & <\Psi^+(2)\Psi(1)>.
\end{eqnarray}
Here $<>$ is the average value with the unknown statistical {\it nonequilibrium}
operator $\rho$ and $1$ denotes the cumulative variables
$({\bf r_1},s_1,t_1...)$ of space, spin, time etc.
The equation of motion for the correlation functions are given
in the form of the Kadanoff-Baym equation
\protect\cite{KB62,Kel64,SL95}
\begin{eqnarray}
  -&i& \left ( G_0^{-1} G^<-G^< G_0^{-1} \right )=
  i \left ( G^R \Sigma^<-\Sigma^< G^A \right ) 
\nonumber\\
&&-i
  \left ( \Sigma^R G^<-G^< \Sigma^A \right )\label{kb}
\end{eqnarray}
where the retarded and advanced functions are introduced as
$A^{R}(1,2)=-i \Theta (t_1-t_2) [A^>\pm A^<]$ and
$A^{A}(1,2)=i \Theta (t_2-t_1) [A^>\pm A^<]$.
Here operator notation is employed where products are understood as integrations
over intermediate variables (time and space) and the upper/lower sign stands for Fermions/Bosons respectively. The Hartree-
Fock drift term
reads
\begin{equation}
  G_0^{-1}(1{1'})=\left (i \hbar {\partial \over \partial t_1}\!+\!
      {\hbar^2 \over 2 m}\nabla_{\bf x_1}^2 \!-\! \Sigma_{HF}(1{1'}) \right ) \delta
  (1\!-\!{1'}) \label{hf}
\end{equation}
with the Hartree Fock self energy
\begin{eqnarray} \label{HF}
&&\Sigma_{HF}(1,1') \nonumber\\
&&= \left ( \mp \delta ({\bf r_1 - r_1'}) \int d{\bf r_2} V({\bf r_1 -r_2})G^<
({\bf r_2}t_1'{\bf r_2}t_1) \right . \nonumber \\
&& \left . + V({\bf r_1 - r_1'})G^< ({\bf r_1}t_1{\bf r_1'}t_1') \right ) \delta (t_1 -t_1')
\end{eqnarray}
where $G({\bf r_2 },t_1,{\bf r_2},t_1)=n({\bf r_2},t_1)$ is the density.

\subsection{Gauge invariance}

In order to get an unambiguous
way of constructing approximations we have to formulate our theory in gauge invariant way.
This can be done following a procedure known from field theory
\protect\cite{I80}. This method has been applied to high field problems in
\protect\cite{BJ91}.
With the help of the Fourier transform of an arbitrary function
G(x,X) over the relative coordinates $x=({\bf r_2}-{\bf r_1},t_2 -
t_1)=({\bf r},\tau)$
with the center of mass coordinates $X=(({\bf r_2}+{\bf r_1})/2,(t_2 +
t_1)/2)=({\bf R},t) $
one can introduce a gauge-invariant Fourier-transform of the difference
coordinates $x$
\be\label{Fouriertrafo}
&&{\bar G}(k,X) = \int dx G(xX)
\nonumber\\&&\times
{\rm exp}
\!\!\left \{ \!\!\frac{i}{\hbar} \int\limits_{-\frac{1}{2}}^{\frac{1}{2}}
d\lambda  x_{\mu}[k^{\mu}\!+\!\frac{e}{c}A^{\mu}(X\!+\!\lambda
x)] \!\!\right \} .
\ee
For constant electric fields, which will be of interest in the
following, one obtains
a generalized Fourier-transform
\[
{\bar G}(k,X) = \int dx\;
{\rm e}^{\frac{i}{\hbar}[x_\mu
k^\mu+e{\bf rE}t]} G(x,X),
\]
where the $\chi$ function was chosen in such a way
that the scalar potential is zero $A^\mu =(0,-c {\bf E}t)$.
Therefore, we have the following rule in formulating the kinetic theory
gauge-invariantly
\begin{enumerate}\label{gaugerule}
\item{Fourier transformation of the 4-dimensional difference-variable x to canonical momentum p.}
\item{Shifting the momentum  to kinematic momentum
according to ${\bf p}={\bf k}-e{\bf E}t$.}
\item{The gauge invariant functions $\bar {G}$ are given by
\be\label{gauge invariant}
&&G({\bf p},t)=G({\bf k}-e{\bf E}t,t)={\bar G}({\bf k},t)\nonumber\\
&&={\bar G}({\bf p}+e{\bf E}t,t).
\ee
}
\end{enumerate}
We shall make use of these rules in the following sections.
In \cite{Mor94} this procedure has been generalized for two - particle
Greens functions and leads to the field - dependent Bethe -
Salpeter - equation.

\subsection{Equation for Wigner distribution}

In the relative and center of mass coordinates
the time diagonal part of (\ref{kb}) reads
\protect\cite{JW84}

\begin{eqnarray}\label{timediagonal}
\frac{\partial}{\partial t}
&&f ({\bf p},t) =
\nonumber\\
&&
\int\limits_0^{t-t_0} d\tau\left[ \left \{ G^>({\bf p},t-\frac{\tau}{2},-\tau),\Sigma^<({\bf
      p},t-\frac{\tau}{2},\tau) \right \} \right .
\nonumber\\
&&-
\left . \left\{ G^<({\bf p},t-\frac{\tau}{2},-\tau),\Sigma^>({\bf p},t-\frac{\tau}{2},\tau) \right \} \right]\nonumber\\
&&
\end{eqnarray}
Here $f({\bf p},t)=G^{<}({\bf p,R},t,\tau=0)$ 
denotes the Wigner distribution function and we suppress the center of
mass coordinates. $\{, \}$ is the
anti-commutator understood that the $\tau$ variable at the first place
comes with a minus
sign respectively.
This equation is exact in time, but according to the assumed slowly varying space dependence
we have used gradient expansion for space variables and dropped all R-dependence for
simplicity. This criterion is discussed in the last section (\ref{space1}).
With the help of the gauge invariant formulation of Green's function
(\protect\ref{gaugerule}), we can write the kinetic equation
(\protect\ref{timediagonal}) finally in the following gauge-invariant form
\begin{eqnarray}\label{tdinv}
&&\frac{\partial}{\partial t}
f({\bf k},t)+e{\bf E}\nabla_{\bf k} f({\bf k},t)
=\int\limits_0^{t-t_0} d\tau
\nonumber\\
&&\left[\!
\left \{\! G^>({\bf k}\!-\!\frac{e{\bf E}}{2}\tau,\tau,t\!-\!\frac{\tau}{2}),
\Sigma^<({\bf k}\!-\!\frac{e{\bf
    E}}{2}\tau,-\tau,t\!-\!\frac{\tau}{2}) \!\right \}_+\right .\nonumber\\
&&\!-\!\left .
\left\{ \! G^<({\bf k}\!-\!\frac{e{\bf E}}{2}\tau,\tau,t\!-\!\frac{\tau}{2}),
\Sigma^>({\bf k}\!-\!\frac{e{\bf E}}{2}\tau,-\tau,t\!-\!\frac{\tau}{2}) \right \}_+ \right]. \nonumber\\
&&
\end{eqnarray}
This kinetic equation is exact in time convolutions. This is necessary because
gradient expansions in time are connected with linearization in electric fields and consequently
fail \protect\cite{m87}. The gradient approximation in space has been applied assuming slow varying processes in space.
This corresponds to the limit of a weakly coupled plasma, which we employed already in Section \ref{Hydrodyn}.
Please remind that due to Coulomb gauge we do not have space inhomogeneity by the electric field.

\subsection{Spectral function}

The spectral properties of the system are described by the
Dyson equation for the retarded Green function.
A free particle in a uniform electric field,
where the field is represented by a vector potential
${\bf E}(t)=-\frac 1 c \stackrel{.}{{\bf A}}(t)$ leads to the following equation
\begin{equation}\label{free}
\left [i\hbar\frac{\partial}{\partial t}-\epsilon({\bf p}-\frac{e}{c}{\bf A}(t)) \right ] G^R_0({\bf p},tt')=
\delta (t-t').
\end{equation}
This equation is easily integrated \protect\cite{jau91,kdw87}
\begin{equation}\label{ret}
G^R_0({\bf p},tt')=-i\Theta(t\!-\!t')\exp{\!\left [{i \over \hbar}\!\int\limits_t^{t'}\!\!du\,
\epsilon({\bf p}\!-\!{e \over c}{\bf A}(u)) \right ]}.
\end{equation}
For free particles and parabolic dispersions,
the gauge invariant spectral function \protect\cite{jau91,kdw87}
follows
\begin{eqnarray}\label{spectrall}
A_0({\bf k},\omega)&=&2\int\limits_0^{\infty}d\tau {\rm cos} \left
( \omega\tau- \frac{k^2}{2m\hbar}\tau-
\frac{e^2E^2}{24m\hbar} \tau^3 \right ) \nonumber \\
&=& \frac{2\pi}{\epsilon_E} Ai \left
(\frac{k^2/2m-\hbar\omega}{\epsilon_E} \right )
\end{eqnarray}
where Ai(x) is the Airy function \protect\cite{a84} and $\epsilon_E=(\hbar^2e^2E^2/8m)^{1/3}$.
It is instructive to verify that (\protect\ref{spectrall}) satisfies the
frequency sum rule $\int d \omega A_0(\omega)=2 \pi$.
The interaction-free but field-dependent
retarded Green's function $G_o^R $ can be obtained from the
interaction-free
and field-free
Green's function by a simple Airy transformation \cite{Moa93}. This is an expression of the fact that
the solutions of the Schr\"odinger equation with constant electric
field are Airy-functions.
The retarded functions can therefore be diagonalized
within those eigen-solutions\protect\cite{BKF89,BJ91}. It can be shown that (\ref{spectrall}) remains valid even within a
quasiparticle picture \cite{Moa93},
where we have to replace simply the free dispersion $k^2/2m$ by the quasiparticle energy $\epsilon_k$.

\subsection{The Problem of the ansatz}
In order to close the kinetic equation
(\protect\ref{timediagonal}), it is
necessary to know the relation between $G^>$ and $G^<$. This problem is known
as an ansatz and must be constructed consistently with the required approximation
of self-energy.
Assuming the conventional KB ansatz \cite{KB62} we have a relation
between the two time Green functions and the distribution function
\beq\label{g}
G^<({\bf k},\omega,{\bf r},t)&=& A({\bf k},\omega,{\bf r},t)\; f({\bf k,r},t)\nonumber\\
G^>({\bf k},\omega,{\bf r},t)&=& A({\bf k},\omega,{\bf r},t)\; (1\mp
f({\bf k,r},t)).
\eeq
This is quite good as long as the quasi-particle picture holds and no memory
effects play any role. As we shall see, the
formulation of kinetic equations with high fields is basically connected with
a careful formulation of retardation times. Therefore, the simple
ansatz,
called {\it KB ansatz} fails.

Another obscure discrepancy is the fact that with the old ansatz,
one has some
minor differences in the resulting collision integrals
compared with
the results from the density operator technique. With the old
ansatz, one gets just
one half of all retardation times in the various time arguments \protect\cite{JW84,jau91}.
This annoying
discrepancy remained obscure until the work of Lipavsky, {\it et al.}
\protect\cite{LSV86} where an expression is given
for the $G^<$ function in terms of expansion after various times.
We can write in Wigner coordinates
\begin{eqnarray}\label{gk}
G^<({\bf p},t,\tau) &=&f({\bf p},t-\frac{|\tau|}{2})A({\bf p},\tau,t).
\end{eqnarray}
This generalized- Kadanoff- Baym (GKB) - ansatz is an exact relation as long as the selfenergy is taken in Hartree- Fock
approximation.
Together with the requirement of gauge invariance of Sec.~\protect\ref{gaugerule}
and using the quasiparticle spectral function
(\ref{spectrall}) with quasiparticle energies $\epsilon_k$
instead of $k^2/2m$, the GKB ansatz finally reads
\be\label{Lipavsky}
&&G^<({\bf k},\tau,{\bf r},t)=\exp{\left \{-\frac{i}{\hbar}
    \left(\epsilon_k \tau + \frac{e^2E^2}{24m}\tau^3\right)\right \}}
\nonumber\\
&&\times
f \left( {\bf k}-\frac{e{\bf E}|\tau|}{2},{\bf r},t-\frac{|\tau|}{2} \right).
\ee
In order to get more physical insight into this ansatz one
transforms into the frequency representation
\beq
&&G^<({\bf k},\omega,{\bf r},t)= 2\int\limits_0^{\infty}d\tau f({\bf
  k}-\frac{e {\bf E}\tau}{2},t-\frac{\tau}{2})
\nonumber\\&&
\times {\rm cos}
\left
(  \omega\tau- \epsilon({\bf k,r},t){\tau \over \hbar}-
\frac{e^2E^2}{24m\hbar} \tau^3 \right )
.
\eeq
Neglecting the retardation in $f$ one recovers the ordinary ansatz
(\ref{g}) with the spectral function (\ref{spectrall}).
The generalized ansatz takes into account
history by an additional memory. This ansatz is superior to the Kadanoff-Baym ansatz in the
case of high external fields in several respects \cite{MJ93}: (i) it has the correct spectral properties, (ii)
it is gauge invariant, (iii) it preserves causality, (iv) the
quantum kinetic equations derived with Eq.(\ref{kinetic}) coincide
with those obtained with the density matrix
technique \cite{L65,L69,JW84}, and (v)
it reproduces the Debye-Onsager relaxation effect \cite{MK92} .

Other choices of ansatz can be appropriate for other physical
situations. For a more detailed discussion see \cite{MLS00}.

\subsection{Kinetic equation in dynamically screened approximation}

For Coulomb interaction it is unavoidable to consider screening if one
does not want to obtain long range or short wave vector divergences.
To obtain an explicit form for the kinetic equation we have to
determine the selfenergy $\Sigma^{>,<}$.
The dynamically screened approximation is given by expressing the self
energy by a sum of all ring diagrams. The resulting kinetic equation
is the quantum Lenard - Balescu equation, which has been derived for high fields in \cite{Moa93}. We give this approximation in exact time convolutions. The selfenergy is given in terms of the dynamical potential ${\cal V}$
\beq
&&\Sigma^<_a({\bf k},t,t')=\int {d {\bf q}\over (2\pi\hbar)^3} {\cal V}^<_{aa}({\bf q},t,t') G_a^<({\bf k-q},t,t')\label{53}
\nonumber\\
&&
\eeq
where the dynamical potential is expressed within Coulomb potentials $V_{ab}({\bf q})$
\beq
{\cal V}^<_{aa}({\bf q},t,t')=\sum\limits_{dc}V_{ad}({\bf q}) {\cal L}^<_{dc}({\bf q},t,t') V_{ca}({\bf q})
\eeq
via the density-density fluctuation
\beq
&&{\cal L}^<_{ab}({\bf q},t,t')=\delta_{ab} \int d{\bar t} d{\bar {\bar t}}\nonumber\\
&&\times
\left ({\cal E}^{r} \right)^{-1} ({\bf q},t,{\bar t})  L_{aa}^<({\bf q},{\bar t},{\bar {\bar t}})
\left ({\cal E}^{a}\right )^{-1}({\bf q},{\bar {\bar t}},t').\label{densityf}
\eeq
Here $L$ is the free density fluctuation
\beq
&&L^<_{aa}({\bf q},t,t')=\int {d {\bf p}\over (2\pi\hbar)^3}G_a^<({\bf p},t,t') G_a^>({\bf p-q},t',t)\label{fluc}
\nonumber\\&&
\eeq
and ${\cal E}^{r/a}$ the retarded/ advanced dielectric function
\beq
&&{\cal E}^{r/a}({\bf q},t,t')=\delta(t-t')\pm i \Theta[\pm(t-t')] \sum\limits_b V_{bb}({\bf q})\nonumber\\
&& \times(L^>({\bf q},t,t')-L^<({\bf q},t,t')).\label{57}
\eeq
One easily convince oneself that this set of equations (\ref{53}-\ref{57}) is
gauge invariant.

We can directly introduce this set of equations into the equation for the Wigner function (\ref{tdinv}) and obtain after some algebra for the in-scattering part of the collision integral
\beq
&&I^{\rm in}_a({\bf k},t)=2 \sum\limits_b \int {d {\bf q}\over
  (2\pi\hbar)^3} V_{ab}^2({\bf q}) \int\limits_{0}^{\infty}d \tau \int
{d \omega \over 2 \pi}
\nonumber\\
&&\times \cos \left [(\epsilon^a_{k-q}-\epsilon_k^a-\omega) \tau+{e_a {\bf E q} \tau^2 \over 2 m_a} \right ]\nonumber\\
&&\times
f_a({\bf k\!-\!q}\!-\!e_a {\bf E} \tau,t\!-\!\tau) (1\!-\!f_a({\bf k}\!-\!e_a {\bf
  E}\tau,t\!-\!\tau))
\nonumber\\
&&\times
{L^<_{bb}({\bf q},\omega, t-\frac 1 2 \tau)\over \left | {\cal E}({\bf
      q},\omega,t-\frac 1 2 \tau) \right |^2} 
\label{eqe}
\eeq
with the free density fluctuation (\ref{fluc})
\beq
&&L_{bb}^<({\bf q},\omega,t)=-2 \int {d {\bf p}\over (2\pi\hbar)^3}
\int\limits_{0}^{\infty} d\tau 
\nonumber\\
&&\times
\cos \left [ (\omega-\epsilon_p^b+\epsilon_{p+q}^b)\tau +{e_b {\bf E
      q} \tau^2 \over 2 m_b}\right ] 
\nonumber\\
&&\times
f_b({\bf p+q},t-\frac 1 2 \tau) (1-f_b({\bf p},t-\frac 1 2 \tau)).
\eeq
The out-scattering term $I^{\rm out}$ is given by $f\leftrightarrow 1-f$.
Here we used the ansatz (\ref{Lipavsky}) and have employed the approximation $t\pm \frac 1 2
\tau \approx t$ in
the density fluctuation (\ref{densityf})  which corresponds to a gradient approximation in times for
the density fluctuations. Since the center of mass time dependence is
carried only by the distribution functions in (\ref{densityf}), this
approximation is exact in the quasistationary case which we investigate
in the next section. All internal time integrations  remain exact.
Of course, for time dependent phenomenae we have to question this approximation.

Eq. (\ref{eqe}) represents the field dependent Lenard-Balescu kinetic
equation \cite{Moa93} which was here slightly rewritten and which form
will turn out to be very convenient for the later analytical integration.
Other standard approximations like the T-matrix \cite{Mor94} approximation resulting into a field dependent Bethe-Salpeter
equation can be given.

\subsubsection{Kinetic equation in statically screened approximation}

Using the static approximation for the dielectric function ${\cal
E}({\bf q},0,t)$
in (\ref{eqe}), the kinetic equation for statically screened
Coulomb potentials in high
electric fields appears \protect\cite{MK92,JW84,Moa93}
\begin{eqnarray}\label{kinetic}
&&\frac{ \partial}{\partial T}  f_a + e {\bf E} {\pa {\bf k_a}} f_a =
\sum_b  I_{ab} \nonumber \\
&&I_{ab} =
\!\frac{2 (2 s_b+1)}{\hbar^2}\!\int \!\! \frac{ d {\bf k'_a} d {\bf k_b} d {\bf k'_b}
}{(2\pi\hbar)^6}
\delta \left({\bf k_a}\!+\!{\bf k_b}\!-\! {\bf k'_a}\!-\!{\bf k'_b}
\right)
\nonumber\\
&&\times\left\{
f_{a'} f_{b'}(1-{f}_a)(1-{f}_b)-{f}_a {f}_b (1-f_{a'})(1-f_{b'})
\right\}
\nonumber\\
&&\times V_{s}^2({\bf k_a}\!-\! {\bf k'_a})
\int\limits_0^{\infty} d\tau
\cos
\left\{
(\epsilon_a+\epsilon_b-{\epsilon'_a}-{\epsilon'_b}){\tau \over \hbar} \right .
\nonumber\\&&
- \left .
\frac{{\bf E}\tau^2}{2 \hbar}
\left(
\frac{e_a{\bf k_a}}{m_a}+ \frac{e_b{\bf k_b}}{m_b}-
\frac{e_a {\bf k'_a}}{m_a}-\frac{e_b {\bf k'_b}}{m_b}
\right)
\right\}\nonumber\\
&&
\end{eqnarray}
with $f_b=f_b(k_b-e_bE\tau,T-\tau)$. 
The potential is the static Debye one
\beq\label{pot}
V_s(p)={4 \pi e_a e_b \hbar^2 \over p^2 +\hbar^2 \kappa^2}
\eeq
and the static screening length $\kappa$ is given by
\beq\label{screen}
\kappa^2=\sum\limits_c {4 \pi e_c^2 n_c \over T_c}
\eeq
in the equilibrium and nondegenerated limit.
Here $T_c$ is the temperature of specie $c$, charge $e_c$, spin $s_c$
and mass $m_c$ respectively.

If we had used the conventional Kadanoff and Baym ansatz (\ref{g})  we would have obtained a factor $1/2$
in different retardations \cite{JW84}. This would lead to no relaxation effect at all \cite{MK92}.
Furthermore it is assumed, that no charge or mass transfer will occur during
the
collision. Otherwise one would obtain an additional term in the
$\cos$ - function
proportional to $\tau^{3}$.

Two modifications of the usual Boltzmann collision integral can be deduced
from (\ref{kinetic}):
({\bf i})A broadening of the $\delta$-distribution function of
the energy conservation
and an additional retardation in the center-of-mass times of the
distribution functions. This is known as collisional broadening and is
a result of the finite collision duration
\protect\cite{SDKW86}. This effect can be observed even if no
external field is applied. It is interesting to remark that this
collisional broadening ensures the conservation of the total
energy \protect\cite{M94}.
If this effect is neglected one obtains the Boltzmann equation for
the field free case.
({\bf ii})
The electric field modifies the broadened $\delta$- distribution
function
considerably by a term proportional to $\tau^2$. This broadening
vanishes for identical charge to mass ratios of colliding
particles. At the same time
the momentum of the distribution function becomes retarded by the
electric field. This effect is sometimes called
intra-collisional-field effect.

\section{Field effects on current}

We are now interested in corrections
to the particle flux, and therefore obtain from (\ref{kinetic}) the balance equation for the momentum
\begin{equation}\label{p}
\frac{\partial}{\partial t} <{\bf k_a}>-n_ae_a{\bf E} =
\sum_b<{\bf k_a}I_B^{ab}>.
\end{equation}

Here we search for the relaxation field (\ref{ek1})
 which will be represented as renormalization of the external field ${\bf E}$
similar to the Debye-Onsager-Relaxation field in the theory of electrolyte
transport
\protect\cite{f53,FEK71,KKE66}. This effect is a result of the
deformation of the two-particle correlation function by an applied electric
field.

To proceed we assume some important restrictions on the
distribution functions. First, we assume a nondegenerate
situation, such that the Pauli blocking effects can be neglected.
Second, to calculate the current for a
quasistationary plasma we choose Maxwellian distributions analog to (\ref{Maxwell})
\beq
f_i(p)={n_i \over 2 s_i+1} \lambda_i^3 \exp{\left \{-{p^2 \over 2 m_i
      T_i}\right \}}
\label{max}
\eeq
with the thermal wave length $\lambda_i^2=2 \pi \hbar^2/(m_i T_i)$,
the spin $s_i$ and
the partial temperature $T_i$ for species $i$ which can be quite different
e.g. in a two - component system.

\subsection{Statically screened result}\label{stat}

Before we present the result for the dynamically screened
approximation we want to give the static result.
The momentum conservation in
(\ref{kinetic}) can be carried out and we get for the relaxation field
\beq
&&n_a e_a {\delta E \over E} {\bf E}=
-\sum\limits_b
 (2 s_a\!+\!1)(2 s_b\!+\!1)\nonumber\\
&&\times \frac{2}{\hbar^2}\int \frac{d {\bf q}d {\bf Q}d {\bf
k}}{(2\pi\hbar)^9} V^2_s(q)
\int\limits_0^\infty d\tau
({\bf k}+e_a {\bf E} \tau) \nonumber\\
&&\times\cos \!\left[\!
 \left (\!\frac{q^2}{2m_a \hbar}\!+\!\frac{{\bf k}{\bf q}}{m_a \hbar}\!-\!
\frac{{\bf q}{\bf Q}}{m_b \hbar}\right)\! \tau \!-\! \frac{{\bf E}{\bf q}}{2 \hbar}\!
\left[ \frac{e_b}{m_b}\!-\!\frac{e_a}{m_a} \right]\! \tau^{2}
\right]
\nonumber\\
&&\times\left\{f_a({\bf k}) f_b \left({\bf Q}+\frac{\bf q}{2}\right)-
f_a({\bf k}+{\bf q}) f_b \left({\bf Q}-\frac{\bf q}{2}\right)
\right\}
\eeq
where we have shifted the retardation into the distribution
functions.
The second part of the distribution functions can be transformed into
the first one by putting ${\bf k+q } \rightarrow {\bf k}$ and ${\bf q} \rightarrow {- \, \bf q}$
with the result
\beq
&&n_a e_a {\delta E \over E} {\bf E}=
\sum\limits_b \frac{2 s_a s_b}{\hbar^2 (2\pi\hbar)^9}\!\int \! d {\bf
k} d {\bf q}d {\bf Q} \, f_b({\bf Q}) \,f_a({\bf k}) \nonumber\\
&\times&V^2(q) \,{\bf q}\int\limits_0^\infty d\tau
\cos
\left[
 \left(
 -\frac{q^2}{2 \mu \hbar}-\frac{{\bf k}{\bf q}}{m_a \hbar}+
\frac{{\bf q}{\bf Q}}{m_b \hbar}
\right)
\tau \right .
\nonumber\\&&
+ \left .\frac{{\bf E}{\bf q}}{2 \hbar}
\left[
\frac{e_b}{m_b}-\frac{e_a}{m_a}
\right]
\tau^{2}
\right]
\eeq
with the reduced mass $\mu^{-1}=1/m_a+1/m_b$. The angular
integrations can be carried out trivially and we get
\beq\label{Ib}
&&n_a e_a {\delta E \over E} {\bf E}={{\bf E} \over E} \sum\limits_b I_1\nonumber \\
&&I_1=\frac{1}{\hbar^{11} 4 \pi^6}\int d q q^3 V^2(q)\!\!
\int\limits_0^\infty\! d\tau
\, {\rm js}\left(
\frac{E q}{2\hbar}
\left[ \frac{e_b}{m_b}\!-\!\frac{e_a}{m_a} \right] \tau^{2}
\right) \nonumber\\
&&\times \sin({q^2 \tau \over 2 \mu \hbar}) I_2[a] I_2[b]\nonumber\\
&&
\eeq
with ${\rm js}(x)=(x \cos x-\sin x)/x^2$. The two integrals over the
distribution functions $I_2$ can be done with the result
\beq
I_2[a]&=&{\hbar m_a (2 s_a+1) \over q \tau} \int\limits_0^{\infty} d k k \,f_a(k)\,
\sin({k q \tau \over m_a \hbar})\nonumber\\
&=&2 \hbar^3 n_a \pi^2 {\rm e}^{-{q^2 \tau^2 T_a\over 2 \hbar^2
m_a}}
\eeq
and correspondingly $I_2[b]$.
We now introduce the new variables
\beq
q&=&2\, y \, \sqrt{\mu T_{ab}}\nonumber\\
t&=&{2 T_{ab} \tau \over \hbar}\nonumber\\
T_{ab}&=&\frac 1 2 \left ({m_b \over m_a+m_b} T_a+{m_a \over m_a+m_b}
T_b \right )\nonumber\\
e&=& {\hbar \sqrt{\mu} E \over 4 T^{3/2}_{ab}}
\left[ \frac{e_b}{m_b}-\frac{e_a}{m_a} \right]
\eeq
and obtain
\beq
&&I_1={8 n_a n_b \mu^2 T_{ab} \over \pi^2 \hbar^4}
\int\limits_0^{\infty} dy y^3 V^2(2 y \sqrt{\mu T_{ab}})
\nonumber\\&&
\int\limits_0^{\infty} dt \,{\rm js}(y t^2 e) \sin(y^2 t) {\rm e}^{-y^2
t^2}.\nonumber\\
\eeq
Using the screened Debye potential (\ref{pot})
we finally obtain
\beq\label{integral}
I_1&=&{8 n_a n_b e_a^2 e_b^2 \over T_{ab}} I_3\nonumber\\
I_3&=&\int\limits_0^{\infty} dz {z^3 \over (z^2+1)^2}
\int\limits_0^{\infty} dl \, {\rm js}(x z l^2) {\sin(z^2 l \zeta ) \over \zeta}
{\rm e}^{-z^2 l^2}.
\nonumber\\&&
\eeq
Therein we used $y=z \zeta$ and $l=t \zeta$ with the quantum parameter
\beq\label{b}
\zeta^2={\hbar^2 \kappa^2 \over 4 \mu T_{ab}}
\eeq
and the classical field parameter
\beq\label{x}
x=\frac e \zeta={E \over 2 T_{ab} \kappa}\!\left (\!{m_a \over m_a+m_b}
e_b\!-\!{m_b \over m_a+m_b}
e_a \! \right ).
\eeq
With this form (\ref{integral}) we have given an extremely useful
representation because the field effects, contained in
$x$, are separated from the quantum effects, which are contained in
$\zeta$.
The integral in  (\ref{integral}) can be performed analytically in the
classical limit $\zeta\rightarrow 0$. For the more general quantum
case with arbitrary $\zeta$ the linear and cubic field effect can be
given analytically and are discussed in \cite{Ma97}. We will not
discuss them here.

Performing the classical limit $\zeta\rightarrow 0$ one obtains from (\ref{integral}) that \cite{O97,Ma97}
\begin{eqnarray}
I_{3c}&=&-{\pi x \over 24} F(|x|)\nonumber\\
F(x)&=&-\frac{3}{x^2}\left [ 3- x +{1 \over 1+x} -\frac 4 x {\rm
ln}(1+x) \right ]\label{classical}.
\end{eqnarray}
Introducing the classical result (\ref{classical}) into
(\ref{integral}) we find from (\ref{Ib}) and (\ref{p}) the following relaxation field
\begin{equation}\label{pe}
\frac{\partial}{\partial t} <{\bf k_a}>-n_ae_a{\bf E}\left (1+{\delta
E_a\over E}\right) =n_ae_a{\bf J} \, R(E)
\end{equation}
with
\beq\label{form}
{\delta E_a \over E}&=&-{e_a \pi \over 6 \kappa} \sum\limits_b
{4 n_b e_b^2\over \mu_{ab}}{{e_b\over m_b}-{e_a\over
m_a}\over \left ({T_b\over m_b}+{T_a\over
m_a}\right )^2} F(|x|)
\eeq
and $x$ from (\ref{x}).
We see that for a plasma consisting of particles with equal charge to
mass ratios,
no relaxation field appears. The link to the known
Debye- Onsager relaxation effect can be found if we assume that we
have a plasma consisting of electrons ($m_e\,,\, e_e=e$) and
ions with charge $e_i=e Z$ and temperatures $T_e=T_i=T$.
Then (\ref{form}) reduces to
\beq\label{form1}
{\delta E_a\over E}&=&-{\kappa e_a^2\over 6 T} {Z
  (1\!+\!\frac{m_e}{m_i} Z) \over (1\!+\!Z)(1\!+\!\frac {m_e}{m_i})} F\!\left ( {e E
    \over T \kappa} {Z (1\!+\!\frac{m_e}{m_i} Z) \over 1\!+\!\frac
    {m_e}{m_i}}\right )
\nonumber\\
&=&- \frac { e^2 \kappa_e}{6 \, T} 
\left\{
\matrix{ \frac 1 2+o(E)\cr\cr
{3 \kappa T\over 2 e E}+o(1/E)^2}
\right . {\rm for}\, Z=1
.
\eeq
This formula together with the general form (\ref{form}) is the main result of
this chapter. It gives the classical relaxation effect for statically screened approximation up to any field
strength and represents a result beyond linear response. We see that
in the case of single charged heavy ions the Debye result
(\ref{debye}) is underestimated by a factor of two.

\subsection{Dynamically screened result}

The calculation of the current with the collision integral for
dynamically screened potentials (\ref{eqe}) can be performed
analytically as well. For the quasistationary condition we can
calculate the frequency integral in (\ref{eqe}) analytically using the
identity \cite{K75} for the classical limit $o(\hbar)$
\beq
&&\int \!\!{d \omega \over 2 \pi} {H(\omega) \over \omega} {\rm Im}
{\cal E}^{-1}(q,\omega)\!=\!\frac{H(0)}{2} {\rm Re} \left ( \!1\!-\!{1
    \over {\cal E}(q,0)}
\!\right ) \label{a13}
\nonumber\\
&&
\eeq
where we set $H(\omega)=\omega/{\rm Im} {\cal E}$ and which
relation is proven in appendix \ref{a1}. We will employ only
classical screening. The quantum result for screening is more involved and not yet analytically integrable.

Observing that for the dielectric function (\ref{57}) together with
(\ref{max}) holds
\beq
\lim\limits_{\omega \to 0} {\omega \over {\rm Im} {\cal E}(q,\omega)}={ q^3\over \sqrt{\pi}\hbar^3} \left (\sum \limits_b {\kappa_b^2 \over v_b}\right )^{-1}
\label{er}
\eeq
with the partial screening length $\kappa_b^2=4 \pi e_b^2 n_b/T_b$ and
the partial thermal velocity $v_b^2=2 T_b/m_b$, we obtain for the
current (\ref{Ib}) after similar integrations as in chapter \ref{stat}
instead of (\ref{integral})
\beq\label{integrald}
&&I_1^{\rm dyn}={8 \kappa^2 e_a^2 e_b^2 n_a n_b \sqrt{m_a m_b}\over \sqrt{\pi \mu_{ab} T T_a T_b} \sum\limits_c {\kappa_c^2\over v_c}} I_3^{\rm dyn}\nonumber\\
&&I_3^{\rm dyn}=\int\limits_0^{\infty} d z {z^2\over 1 + z^2}
\int\limits_{-1}^1 d x x \int\limits_0^{\infty} d l d l_1 {\rm
  e}^{-z^2 (l^2+l_1^2)}
\nonumber\\
&&\times\frac 1 \zeta \cos[M_b \zeta l z^2+B z l^2 x] \cos[M_a \zeta l_1 z^2 -A z l_1^2 x].
\eeq
Here we used the same dimensionless variables as in chapter \ref{stat} and the quantum parameter (\ref{b}). Further we abbreviated
$A={e_a E\over \kappa T_a}$,
$B={e_b E\over \kappa T_b}$,
$M_a=\sqrt{2 \mu T \over m_a T_a}$,
$M_b=\sqrt{2 \mu T \over m_b T_b}$.

We like to remark that we neglect any field dependence on the
screening ${\cal E}$ itself here. As presented in \cite{MJ93} a field dependent
screening function can be derived. However, this field dependence
gives rise to a field dependence starting quadratically and will be
not considered in this work.

The classical limit of (\ref{integrald}) can be performed again by $\zeta \rightarrow 0$.
We obtain
\beq\label{i2d}
I_3^{\rm dyn}=\frac 1 2 A M_a I[|A|,|B|]\,\,-\,\, (a \leftrightarrow b)
\eeq
with the remaining 3-dimensional integral
\beq\label{iab}
&&I[A,B]=\int\limits_0^{\infty} d z {z^3 \over z^2+1} \int\limits_{-1}^1
dx {x^2\over A^2 x^2+z^2} \nonumber\\
&&\int\limits_0^{\infty} d l {\rm e}^{-z^2 l^2} \cos(B z l^2 x).
\eeq

\subsubsection{Linear response}

The linear response can be read off directly from (\ref{i2d}) and is given by
$I[0,0]$ of (\ref{iab}). We obtain
\beq
I_3^{\rm dyn}={\pi^{3/2} \over 12} (M_a A-M_b B)
\eeq
and the linear relaxation field (\ref{form}) takes the form
\beq\label{form3}
&&{\delta E^{\rm dyn} \over E}={4 e \pi \kappa \over 3 \sum\limits_c \kappa_c^2 \sqrt{m_c\over T_c}} \sum\limits_b
n_b e_b^2\sqrt{m_a m_b \over T_a T_b} 
\nonumber\\
&&\times\left ({e_a \over T_a^{3/2}\sqrt{m_a}}-{e_b\over T_b^{3/2} \sqrt{m_b}} \right )+o(E).
\eeq
The difference to (\ref{form}) becomes more evident if we consider
again only electrons and ions with equal temperature
\beq\label{form2}
{\delta E^{\rm dyn}\over E}&=&-{\kappa e^2\over 6 T} {2 Z (1+\sqrt{\frac{m_e}{m_i}} Z) \over  (Z+\sqrt{\frac {m_e}{m_i}})}+o(E).
\eeq
The differences to (\ref{form1}) are obvious in the different mass dependence. 
This result overestimates the Debye result by a factor of two.

\subsubsection{Complete classical result}

Now we are able to present a complete field dependence beyond linear response. The integral (\ref{iab}) can be done analytically, which is sketched in appendix \ref{b1}. The result reads
\beq
&&I[A,B]={\pi^{3/2}\over 6} {\cal I}[A,B]\nonumber\\
&&{\cal I}[A,B]={3 \over 2 A^3} \!\left [ \!
{{4A(1\! -\!{\sqrt{1 + B}} }\over B}
\!+\!{{{A^2} \!+\! \log (1 \!-\! {A^2})}\over {{\sqrt{1 + {B\over A}}}}} 
\right .\nonumber\\&&
\left .+2\,\left( {{{\rm ArcTanh}({1\over
            {{\sqrt{1 - {B\over A}}}}}) -
         {\rm ArcTanh}({{{\sqrt{1 + B}}}\over
            {{\sqrt{1 - {B\over A}}}}})}\over
       {{\sqrt{1 - {B\over A}}}}} \right .\right .\nonumber\\
&&+\left . \left .
     {{-{\rm ArcTanh}({1\over {{\sqrt{1 + {B\over A}}}}}) +
         {\rm ArcTanh}({{{\sqrt{1 + B}}}\over
            {{\sqrt{1 + {B\over A}}}}})}\over
       {{\sqrt{1 + {B\over A}}}}} \right)  
\right ].\label{int2}
\eeq
We obtain for (\ref{form3})
\beq\label{form4}
&&{\delta E_a^{\rm dyn} \over E}={4 e_a \pi \kappa \over 3 \sum\limits_c \kappa_c^2 \sqrt{m_c\over T_c}} \sum\limits_b
n_b e_b^2\sqrt{m_a m_b \over T_a T_b} 
\nonumber\\&&\times
\left ({e_a \over T_a^{3/2}\sqrt{m_a}}{\cal I}[A,B]-{e_b\over T_b^{3/2} \sqrt{m_b}} {\cal I}[B,A] \right ).
\eeq
Expanding (\ref{int2}) in powers of $E$ we recover (\ref{form3}). Once more we choose the case of electrons and ions with equal temperature and obtain
\beq\label{form5}
{\delta E^{\rm dyn}\over E}&=&-{\kappa e^2\over 6 \epsilon_0 T} {2
  Z ({\cal I}[A,B]+\sqrt{\frac{m_e}{m_i}} Z \, {\cal I}[B,A]) \over  (Z+\sqrt{\frac {m_e}{m_i}})}.
\eeq
For single charged ions and big mass differences we can further simplify to
\beq\label{form6}
&&{\delta E^{\rm dyn}\over E}=-{\kappa e^2\over 6 T}
{\cal F}[{e E\over \kappa T}]
\nonumber\\
&&=- \frac { e^2 \kappa_e}{6 \, T} 
\left\{
\matrix{ 2+o(E)\cr\cr
{3 \kappa T\over \sqrt{2} e E}+o(1/E)^2}
\right .
\nonumber\\
&&{\cal F}[x]={3 \over {{x^3}}} \,\left( {{2\,\left( -2\,x +
             3\,\left( -1 + {\sqrt{1 + x}} \right)  \right) }
\over {{\sqrt{1 + x}}}} \right .\nonumber\\
&&\left . +
       {\sqrt{2}}\,\left( -{\rm ArcTanh}({1\over
              {{\sqrt{2}}}}) +
          {\rm ArcTanh}({{{\sqrt{1 + x}}}\over {{\sqrt{2}}}})
\right)  
\right .
\nonumber\\&&
\left .
+ {{{x^2} + \log (1 - {x^2})}\over
         {{\sqrt{2}}}} \right)=2+o(E) .
\eeq
This result will be compare with the statically screened result
(\ref{form1}) and the hydrodynamical result (\ref{9}) in section
\ref{dis}. Here we remark already that the Debye result is
twice overestimated here.

\subsection{Thermally averaged dynamically screened result}\label{dis}

We will now give an approximative treatment of the dynamical
screening used in \cite{er79}. This approximation consists into
the replacement of the dynamical screening in the collision
integral (\ref{eqe}) which is $\overline{{\cal E}(\omega,q)^{-2}}$ by $(1+\kappa^2V_{aa}(q)/4\pi)^{-1}$.
This represents a thermal averaging \cite{K75}
of ${\cal E}^{-2}$ which can
be proven easily with the help of appendix \ref{a1}.
We obtain the relaxation effect of (\ref{form}) and (\ref{form1})
but with a different field function $F$
\begin{eqnarray}
F^{\rm dyn}(x)&=&-\frac{3}{x^2}\left [ 2- x -{2 \over x}{\rm
ln}(1+x) \right ]
\nonumber\\
&=&\left\{
\matrix{ 2+o(x)\cr\cr
{3 \over  x}+o(1/x)^2}
\right .
.\label{classical1}
\end{eqnarray}
Therefore the relaxation effect (\ref{form1}) in linear response for single
charged ions takes the form of (\ref{debye}) and twice the
static screened result (\ref{form1}) and half of the dynamical
screened result (\ref{form6}).

\begin{figure}
\centerline{\psfig{file=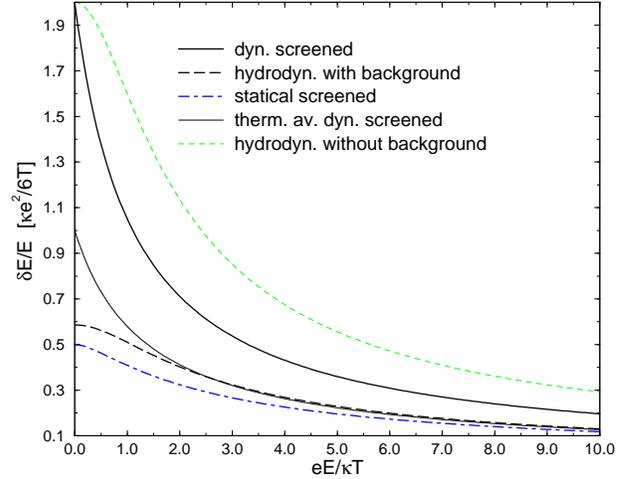,width=8cm,angle=-90}}
\caption{\label{1}The nonlinear Debye-Onsager relaxation effect
  vs. scaled electric field for an electron and single charged ion system. The
  hydrodynamical approximation (\protect\ref{9}) leads to the Onsager
  result (\protect\ref{ons}) for small field strength $2-\sqrt{2}$. 
The statically
  screened result (\protect\ref{form}) or (\protect\ref{form1}) leads
  to half the Debye result (\protect\ref{debye}). The thermally
  averaged approximation of the dynamical screening
  (\protect\ref{classical1}) leads to the Debye result while the full dynamically
  screened approximation (\protect\ref{form6}) leads to twice the
  Debye result. Also the hydrodynamical result (\protect\ref{erg2}) without background
  leads to twice the Debye result. }\label{fig1}
\end{figure}

As we see from figure \ref{fig1} the different approximations lead to very different
results. The statically screened result (\ref{form1}) underestimates
the Debye result by factor 2 which is corrected by the thermal averaged
treatment of the screening. If we calculate instead the complete
dynamical screened result (\ref{form2}) or (\ref{form5}) we
obtain twice the Debye result (\ref{debye}) and the thermal averaged screened
result. However there is a complete different charge dependence. 
We have to observe that the perfectly symmetric treatment of screening
does not reproduce the hydrodynamical result which is the
Onsager result (\ref{ons}) for linear response.
 
\subsection{Asymmetric dynamical screened result}

We want now to proceed and ask under which assumptions the
Onsager result (\ref{ons}) might be reproduced. Following the results we saw from
the hierarchy we have consequently to treat the electrons (specie a) and
ions (all other species) 
asymmetrically. This we will perform in the same spirit as Onsager in
that the ions have to be treated dynamical (as before) but the
electrons are screened statically. 

This means we consider as the potential not the bare Coulomb one but
a statically screened Debye potential for specie a. The
ions (all other species) will
then form the dynamical screening. In comparison with the chapter
before we can perform all steps analogously except two modifications,
eq. (\ref{mod1}) and (\ref{form51}).
First we observe that instead of (\ref{er}) we have now
\beq
\lim\limits_{\omega \to 0} {\omega \over {\rm Im} \,{\cal
    E}(q,\omega)}={ q v_b (q^2+ \hbar^2 \kappa_e^2) \over \sqrt{\pi}\hbar^3\kappa_b^2}
\label{er1}
\eeq
which leads to a replacement of the sum 
\be
\sum\limits_c{\kappa_c^2\over
  v_c}\rightarrow {\kappa^2\over v_i}
\ee 
in the for-factor of (\ref{integrald}) and (\ref{form4}). This leads in the limit of big mass differences to a
for-factor in (\ref{form2}) and (\ref{form5}) respectively
\be
{\rm modification\,I:}\qquad {Z\over 1+Z}.
\label{mod1}
\ee
The second modification is that in (\ref{integrald}) one has to
replace
\be
{z^2\over 1+z^2}&\rightarrow& {z^2\over 1+z^2} {z^2\over
  q+z^2}\nonumber\\
&=&{q\over q-1}{z^2\over q+z^2}-{1\over q-1}{z^2\over
  1+z^2}
\ee
with
\be
q={\kappa_a^2\over \kappa^2}.
\ee
This shows that in the end results (\ref{form4}), Eq. (\ref{form5}) has to be changed
\be
&&{\rm modification \, II:}\nonumber\\
&&{\cal I}[A,B]\rightarrow {\sqrt{q}\over q-1}{\cal I}[{A\over \sqrt{q}},{B\over
  \sqrt{q}}]-{1\over q-1} {\cal I}[A,B].
\label{form51}
\ee
Particularly  we obtain for the linear response result (\ref{form2})
where for electron-ion plasma $q=1/(Z+1)$ and
\beq\label{form21}
{\delta E^{\rm asy}\over E}&=&{\delta E^{\rm dyn}\over E} {Z
  q\over \sqrt{q}+1}\nonumber\\
&=&
-{\kappa e^2\over 3  T}{Z
  q\over \sqrt{q}+1} +o(E)\nonumber\\
\eeq
which agrees with (\ref{erg1}) if we consider that the mobilities are
very different $b_i/b_e\rightarrow0$ in (\ref{erg1a}).

The same result we obtain from the thermally averaged result
(\ref{classical}) since there appears no such function as (\ref{er1})
and therefore the modification I of (\ref{mod1}) does not apply but
solely the modification II of (\ref{form51}). We therefore obtain
(\ref{form1}) but
\be
F^{\rm dyn}_{\rm asy}(x)&=&2 F^{\rm dyn}(x)-\sqrt{2} F^{\rm dyn}(\sqrt{2} x)
\nonumber\\
&=&\left\{
\matrix{ 2-\sqrt{2}+o(x)\cr\cr
{3 \over  2 x}+o(1/x)^2}
\right .
\label{classical2}
\ee
with $F^{\rm dyn}$ of (\ref{classical1}). The linear response leads
then exactly to the same result as from the dynamical screening
(\ref{form21}),
i.e. the Onsager result with the same charge dependence.

The fact that
we reproduce the classical Onsager result with the same charge
dependence can be considered as very satisfactory. The more since we have
seen how many different considerations are possible. 
Please note that 
the special case $Z=1$ could lead occasionally to a seemingly
agreement between different treatments. We think that the charge
dependence incriminates different treatments.

\begin{figure}[h]
\psfig{file=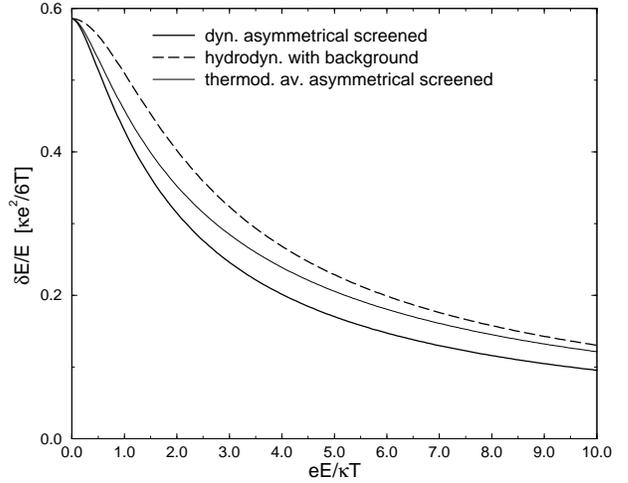,width=8cm,angle=-90}
\caption{The different asymmetric screened approximations for the relaxation effect versus field strength. The hydrodynamical result
(\protect\ref{9}) is compared with the thermally averaged asymmetric
screened result of (\protect\ref{classical2}) as well as the asymmetric
screened one of (\protect\ref{form51}).\label{fig2}}
\end{figure}

In figure \ref{fig2} we see that
the asymmetrical screened result (\ref{form5}) with (\ref{form51})
approaches the hydrodynamical or Onsager result (\ref{ons}) rather well for small fields
while it is too low at high fields. On the other hand the thermally
averaged symmetrical screened result (\ref{classical2}) agrees with
the hydrodynamical approximation (\ref{9}) in the low and high field limit.
Why the hydrodynamical result cannot be reproduced completely within
the kinetic theory remains still a puzzle. Probably the remaining
difference is due to the neglect of the field effect on the screening
itself
\cite{MJ93}.

\section{Range of applicability}

During the derivation of the quantum kinetic equations there has been
assumed the gradient approximations which restricted the spatial
gradients of the system. Here we want to discuss up to which field
strength
this assumption is justified.

The electric field is limited to values $x<<1$ for $x$ from
(\ref{x}). This can be deduced from the expression for the
dynamical screened result (\ref{form6}). The expression has a
remove-able singularity at $x=1$.
Therefore we see a smooth curve. Nevertheless this is the field
strength where something is happening.
For equal masses and temperatures of plasma components this condition translates into
\beq\label{cond1}
E<{\kappa T\over e}.
\eeq
We interpret the occurrence of such singular point that
no thermal distributions are
pertained in the system. Then we have to take into account non-thermal field dependent distributions which have been
employed to study nonlinear conductivity \cite{MSK93,KMSR92,MK93}.

The condition (\ref{cond1}) allows for different physical interpretations. Within the picture of the screening cloud we can
rewrite (\ref{cond1}) into
\beq
e E < m {v_{\rm th}^2 \over r_D}.
\eeq
This means that a particle moving on the radius of the screening cloud $r_D=1/\kappa$ with thermal velocity $v_{\rm
th}^2=T/m$ should not be pulled away by the acting field force. 
We can discuss this limit also via the energy density which can be reached in a system by the applied field. We can
reformulate once more the condition (\ref{cond1}) to find equivalently
\beq
{E^2 \over 4 \pi}<n T.
\eeq
This means that we have essentially non-thermal effects to be expected
if the energy density of the field becomes comparable with the thermal energy density.

The validity criterion (\ref{cond1}) can now be used to check the weak space inhomogeneity which has been assumed during
our calculation. Quasi- equilibrium in charged systems with external fields can only be assumed if the field current is accompanied
by an equivalent diffusion current
\beq
j_{\rm field}=e \mu E n=-j_{\rm diff}=eD{d n \over dx}
\eeq
using the Einstein  condition $\mu=eD/T$ one gets
\beq
eE=T \frac 1 n {d n \over dx}.
\eeq
Combining this elementary consideration with our condition (\ref{cond1})
we obtain a limitation for space gradients
\beq\label{space1}
{d n \over d\,(\kappa \,x)}<n
\eeq
where our treatment of field effects and local equilibrium is applicable.

\section{Summary}

The nonlinear relaxation field of a charged system under the influence
of high electric fields is investigated.
The local equilibrium or hydrodynamical approach starting from the
classical BBGKY hierarchy is compared with the results from the
quantum kinetic equations.

We come to the same conclusion considering the hydrodynamical
approximation or the kinetic theory 
that a perfectly symmetric two component plasma will
lead to a different relaxation effect than the case where we consider
the moving charge asymmetrically from the screening surrounding. In
the hydrodynamic approach this has been achieved by friction with a
background, in the kinetic approach we have realized it due to
asymmetric screening. Within this asymmetric treatment the limit to a one component plasma which would be to
set
the ion charge to $Z=-1$ leads to a non-vanishing finite quantity. 
Oppositely in the
perfectly symmetrical treatment this limit vanishes in that the relaxation
field vanishes as it should. The perfectly symmetrical treatment of
species in the system leads to twice the Debye result different from
the Onsager result in linear response.

Different approximations of the kinetic approaches are compared and
discussed. We found that the symmetrical treatment of species as well
as the asymmetrical treatment leads to the same corresponding results as the
hydrodynamical approach for linear response. But for higher field
strengths there appear minor differences which are probably due to the
neglect of the field dependent screening itself. The thermally
averaged approximation of screening has the advantage to agree for low
and high fields with the hydrodynamical of local equilibrium approach.

\acknowledgements

The author wants to thank warmly Jens Ortner who has contributed the 
hydrodynamical approach with background. Also many comments regarding the
manuscript are gratefully acknowledged. Last but not least
Gerd R{\"o}pke is thanked for stimulating intellectual disagreements
which have provoked this work.

\appendix
\section{Integrals over dielectric functions}\label{a1}

Here we proof a very useful relation, which has been given in \cite{K75}.
Therefore we consider the following integral including the dielectric function
\beq\label{a10}
I&=&\int {d \omega \over 2 \pi} {H(\omega) \over \omega} {\rm Im } \epsilon^{-1}(\omega)\nonumber\\
&=&\int \! {d \omega \over 4 \pi i} \!\left (\!{1\over \omega +i \eta} \!+\!
{1\over \omega -i \eta}\! \right ) H(\omega) (f^-\!-\!f^+)
\eeq
where $f^+=1-1/\epsilon$ and $f^-=(f^+)^*$. In the following we will assume that the function $H(\omega)$ is analytical and vanishes with $\sim \omega^{-2}$ for large $\omega$.
Since $f^{\pm}(\omega)$ has no poles in the lower/upper half plane we have the identity
\beq
\int {d \omega \over 2 \pi i} H(\omega) {f^{\pm}(\omega) \over (\omega \pm i \eta)} =\mp f^{\pm}(0) H(0)\label{a11}
\eeq
and all other combinations of $f^{\pm}$ with the denominator vanish. If we would use the quantum dielectric function $\epsilon$ we would have to add the residue of the poles at the Matsubara frequencies. Because we calculate only with the classical dielectric function we can use (\ref{a11})
With the help of the relation (\ref{a11}) we compute easily for (\ref{a10})
\beq
I=\frac 1 2 H(0) {\rm Re} \left ( 1-{1 \over \epsilon(0)}\right )
\eeq
which proves relation (\ref{a13}).

\section{An Integral}\label{b1}

Here we calculate the integral (\ref{iab}) 
\beq
I[a,b]&=&\int\limits_0^{\infty} d z {z^3 \over z^2+1} \int\limits_{-1}^1
dx {x^2\over a^2 x^2+z^2} \nonumber\\&\times&
\int\limits_0^{\infty} d l {\rm e}^{-z^2 l^2} \cos(b z l^2 x).
\eeq
The variable substitutions $l \rightarrow p$ by $p=\sqrt{z} l$, $x \rightarrow z$ by $z=y x$ and $p \rightarrow e$ by $p \sqrt{x}=l$ leads to 
\beq
&&I[a,b]\nonumber\\&&
=2 \!\int\limits_0^1 \!dx \!\int\limits_0^{\infty} \!dy {y^{5/2} \over y^2 x^2+1}
{x^3 \over a^2+y^2} \!\int \limits_0^{\infty} \!d e {\rm e}^{-y e^2}\cos{(b e^2)}\nonumber\\
&&=\int\limits_0^{\infty} \!dy {y^{1/2} \over a^2 +y^2} \left
  (\!1\!-\!{\log(1+y^2) \over y^2}\! \right )\!\!\int\limits_0^{\infty}\!\! d e
{\rm e}^{-y e^2}\cos{(b e^2)}
\nonumber\\&&
\eeq
where the trivial $x$- integration has been carried out. The variable substitution $e->\rightarrow q$ by $\sqrt{y} e=q$ and $y\rightarrow z$ by $y=1/z$ leads to 
\beq
&&I[a,b]=\frac{1}{a^2} \!\int \limits_0^{\infty}\! dq {\rm e}^{-q^2}\!\!\int\limits_0^{\infty}\!\! dy \cos(b q^2 z) {1\!-\!z^2 \log(1\!+\!{1\over z^2}) \over z^2 \!+\!{1 \over a^2}}.\label{b11}
\nonumber\\&&
\eeq
Now we proceed and use an integral calculated in the next subsection \ref{b2}
\beq
&&\int\limits_{-\infty}^{\infty} dy {\rm e}^{i c y} {1-y^2
  \log(1+{1\over y^2})\over y^2+{1 \over a^2}}=
2 \pi \int\limits_0^1 dx x^2 {{\rm e}^{-c x} \over {1\over a^2}-x^2}
\nonumber\\&&
+a \pi {\rm e}^{-c/a} (1+{\log (1-a^2)\over a^2})\label{b14}
\eeq
to obtain for (\ref{b11})
\beq
I[a,b]&=&{\pi^{3/2} \over 4 a} {1+{\log(1-a^2) \over a^2}\over \sqrt{b/a}}\nonumber\\&&+
{\pi^{3/2}\over 2 a^2} \int\limits_0^1 dx {x^2\over (\frac{1}{a^2}-x^2)\sqrt{1+b x}}.
\eeq
The last integrals is trivial and we end up with (\ref{int2}).

\subsection{Another Integral}\label{b2}

\begin{figure}
\epsfxsize=8cm
\centerline{\epsffile{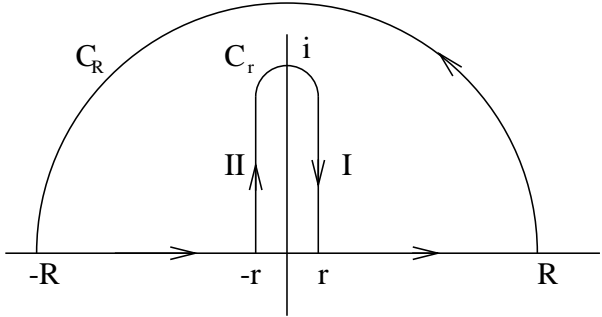}}
\caption{\label{path1}The complex integration path for the integral (\protect\ref{b10}).}
\end{figure}

Our task remains now to solve the integral
\beq
I=\int\limits_{-\infty}^{\infty} dy {\rm e}^{i c y} {1-y^2 \log(1+{1\over y^2})\over y^2+{1 \over a^2}}.\label{b10}
\eeq
Because the complex function $\log(1+1/y^2)$ has a cut from $(0,i)$ we perform the integration along the path as depicted in figure \ref{path1} and write
\beq
&&\int\limits_R^{-r}+\int\limits_r^R+C_R+C_r
+\int\limits_I+\int\limits_{II}
\nonumber\\&&
=2 \pi i \, {\rm Res}\left [{{1-y^2 \log(1+{1\over y^2})\over y^2+{1 \over a^2}},i/a}\right ]\nonumber\\
&&=\pi a {\rm e}^{-c/a} (1+{\log(1-a^2) \over a^2}).\label{b12}
\eeq
It is now easy to proof that in the limit $r\rightarrow 0$ and $R\rightarrow \infty$ the integration parts $C$ vanish. Since the first two parts of (\ref{b12}) represent just the desired integral $I$ we have to calculate
\beq
\int\limits_I+\int\limits_{II}&=&\int\limits_{i+r}^r dy {\rm e}^{i c
  y} {1-y^2 \log(1+{1\over y^2})\over y^2+{1 \over a^2}}
\nonumber\\&&
+\int\limits_{-r}^{i-r} dy {\rm e}^{i c y} {1-y^2 \log(1+{1\over y^2})+2 \pi i\over y^2+{1 \over a^2}}\nonumber\\
&=&-2 \pi \int\limits_0^1 dx {x^2 {\rm e}^{-c x} \over {1\over a^2}-x^2}\label{b13}.
\eeq
Using (\ref{b13}) and (\ref{b12}) we obtain just (\ref{b14}).

\bibliography{kmsr,kmsr1,kmsr2,kmsr3,kmsr4,kmsr5,kmsr6,kmsr7,vac1,spin,delay2,refer,gdr,delay3}

\begin{thebibliography}{10}

\bibitem{HJ96}
H. Haug and A.~P. Jauho, {\em Quantum Kinetics in Transport and Optics of
  Semiconductors} (Springer, Berlin Heidelberg, 1996).

\bibitem{THWS96}
W. Theobald, R. H{\"a}{\ss{}}ner, C. W{\"u}lker, and R. Sauerbrey, Phys. Rev.
  Lett. {\bf 77},  298  (1996).

\bibitem{kker86}
W.~D. Kraeft, D. Kremp, W. Ebeling, and G. R{\"o}pke, {\em Quantum Statistics
  of Charged Particle Systems} (Akademie Verlag, Berlin, 1986).

\bibitem{KES93}
Y. Kluger, J.~M. Eisenberg, and B. Svetitsky, Int. J. Mod. Phys. {\bf 2},  333
  (1993).

\bibitem{k58}
B.~B. Kadomtsev, Zh.Eksp.Teor.Fiz. {\bf 33},  151  (1958), sov. Phys. -JETP
  33,117(1958).

\bibitem{KE72}
Y.~L. Klimontovich and W. Ebeling, Jh. Eksp. Teor. Fiz. {\bf 63},  904  (1972).

\bibitem{e76}
W. Ebeling, Ann. Phys. {\bf 33},  5  (1976).

\bibitem{er79}
W. Ebeling and G. R{\"o}pke, Ann. Phys. (Leipzig) {\bf 36},  429  (1979).

\bibitem{r88}
G. R{\"o}pke, Phys.Rev.A {\bf 38},  3001  (1988).

\bibitem{MK92}
K. Morawetz and D. Kremp, Phys. Lett. A {\bf 173},  317  (1993).

\bibitem{ER98}
A. Esser and G. R{\"o}pke, Phys. Rev. E. {\bf 58},  2446  (1998).

\bibitem{DH23}
P. Debye and E. H{\"u}ckel, Phys. Zeitsch. {\bf 15},  305  (1923).

\bibitem{O27}
L. Onsager, Phys. Zeitsch. {\bf 8},  277  (1927).

\bibitem{f53}
H. Falkenhagen, {\em Elektrolyte} (S. Hirzel Verlag, Leipzig, 1953).

\bibitem{FEK71}
H. Falkenhagen, W. Ebeling, and W.~D. Kraeft,  in {\em Ionic Interaction},
  edited by Petrucci (Academic Press, New York and London, 1971), Chap.~1, p.\
  1.

\bibitem{KKE66}
D. Kremp, D. Kraeft, and W. Ebeling, Ann. Phys. (Leipzig) {\bf 18},  246
  (1966).

\bibitem{Ma97}
K. Morawetz, Contrib. to Plasma Physics {\bf 37},  195  (1997), errata:37,4.

\bibitem{O97}
J. Ortner, Phys. Rev. E {\bf 56},  N5  (1997).

\bibitem{B46}
N.~N. Bogoliubov, J. Phys. (USSR) {\bf 10},  256  (1946), transl. in {\it
  Studies in Statistical Mechanics}, Vol. 1, editors D. de Boer and G. E.
  Uhlenbeck (North-Holland, Amsterdam 1962).

\bibitem{KE62}
Y.~L. Klimontovich and W. Ebeling, Jh. Eksp. Teor. Fiz. {\bf 43},  146  (1962).

\bibitem{D84}
P. Danielewicz, Ann. Phys. (NY) {\bf 152},  239  (1984).

\bibitem{SL94}
V. {\v S}pi{\v c}ka and P. Lipavsk{\'y}, Phys. Rev. Lett {\bf 73},  3439
  (1994).

\bibitem{SL95}
V. {\v S}pi{\v c}ka and P. Lipavsk{\'y}, Phys. Rev. B {\bf 52},  14615  (1995).

\bibitem{M94}
K. Morawetz, Phys. Lett. A {\bf 199},  241  (1995).

\bibitem{lkaw91}
P. Lipavsk{\'y}, F.~S. Khan, F. Abdolsalami, and J.~W. Wilkins, Phys. Rev. B
  {\bf 43},  4885  (1991).

\bibitem{KB62}
L.~P. Kadanoff and G. Baym, {\em Quantum Statistical Mechanics} (Benjamin, New
  York, 1962).

\bibitem{Kel64}
L.~V. Keldysh, Zh.exper.teor.Fiz. {\bf 47},  1515  (1964).

\bibitem{I80}
C. Itzykson and J.~B. Zuber, {\em Quantum field theory} (McGraw-Hill, New York,
  1990).

\bibitem{BJ91}
R. Bertoncini and A.~P. Jauho, Phys.Rev.B {\bf 44},  3655  (1991).

\bibitem{Mor94}
K. Morawetz and G. R{\"o}pke, Zeit. f. Phys. A {\bf 355},  287  (1996).

\bibitem{JW84}
A.~P. Jauho and J.~W. Wilkins, Phys. Rev. B {\bf 29},  1919  (1984).

\bibitem{m87}
G.~D. Mahan, Phys.Rep. {\bf 145},  251  (1987).

\bibitem{jau91}
A.~P. Jauho,  in {\em Quantum Transport in Semiconductors}, edited by D. Ferry
  and C. Jacoboni (Plenum Press, New York, 1991), Chap.~7.

\bibitem{kdw87}
F.~S. Khan, J.~H. Davies, and J.~W. Wilkins, Phys, Rev. B {\bf 36},  2578
  (1987).

\bibitem{a84}
M. Abramowitz and I.~A. Stegun, {\em Pocketbook of mathematical functions}
  (Verlag Harri Deutsch, Frankfurt/Main, 1984).

\bibitem{Moa93}
K. Morawetz, Phys. Rev. E {\bf 50},  4625  (1994).

\bibitem{BKF89}
R. Bertoncini, A.~M. Kriman, and D.~K. Ferry, Phys.Rev.B {\bf 40},  3371
  (1989).

\bibitem{LSV86}
P. Lipavsk{\'y}, V. {\v S}pi{\v c}ka, and B. Velick{\'y}, Phys. Rev. B {\bf
  34},  6933  (1986).

\bibitem{MJ93}
K. Morawetz and A.~P. Jauho, Phys. Rev. E {\bf 50},  474  (1994).

\bibitem{L65}
I.~B. Levinson, Fiz. Tverd. Tela Leningrad {\bf 6},  2113  (1965).

\bibitem{L69}
I.~B. Levinson, Zh. Eksp. Teor. Fiz. {\bf 57},  660  (1969), [Sov. Phys.--JETP
  {\bf 30}, 362 (1970)].

\bibitem{MLS00}
K. Morawetz, P. Lipavsk{\'y}, and V. {\v S}pi{\v c}ka, Phys. Rev. B  (2000),
  sub. cond-mat/0005287.

\bibitem{SDKW86}
S.~K. Sarker, J.~H. Davies, F.~S. Khan, and J.~W. Wilkins, Phys, Rev. B {\bf
  33},  7263  (1986).

\bibitem{K75}
Y.~L. Klimontovich, {\em Kinetic theory of nonideal gases and nonideal plasmas}
  (Academic Press, New York, 1975).

\bibitem{MSK93}
K. Morawetz, M. Schlanges, and D. Kremp, Phys. Rev. E {\bf 48},  2980  (1993).

\bibitem{KMSR92}
D. Kremp, K. Morawetz, M. Schlanges, and V. Rietz, Phys. Rev. E {\bf 47},  635
  (1993).

\bibitem{MK93}
K. Morawetz and D. Kremp, Phys. Fluids B {\bf 1},  225  (1994).

\end{thebibliography}
\bibliographystyle{prsty}

\end{document}